\documentclass[
 twocolumn,
 floatfix,
 a4paper,
 showpacs,
 showkeys,
 nofootinbib,
 reprint
 ]{revtex4-1}

\usepackage{epsfig}
\usepackage{latexsym}
\usepackage{xspace}
\usepackage[colorlinks=true,linktocpage=true,linkcolor=blue,citecolor=blue,allcolors=blue]{hyperref}
\usepackage[utf8]{inputenc}
\usepackage{indentfirst}
\usepackage{enumerate}
\usepackage{color}
\usepackage{todonotes}

\usepackage[caption=false,position=top]{subfig}

\usepackage{amsmath}
\usepackage{amssymb}
\usepackage[english]{babel}
\usepackage{url}

\newcommand{\non}{\nonumber \\}
\newcommand{\Pgce}{P^{\rm gce}}
\newcommand{\Pce}{P^{\rm ce}}
\newcommand{\Bin}{B_{\rm in}}
\newcommand{\Bout}{B_{\rm out}}
\newcommand{\Nin}{N_{\rm in}}
\newcommand{\inout}{\rm in/out}

\newcommand{\eq}[1]{\begin{align} #1 \end{align}}
\newcommand{\mean}[1]{\langle #1 \rangle}

\newcommand{\sNN}{\sqrt{s_{\rm NN}}}

\begin{document}


\title{
Correcting event-by-event fluctuations in heavy-ion collisions for exact global conservation laws with the generalized subensemble acceptance method
}

\author{Volodymyr Vovchenko}
\affiliation{Nuclear Science Division, Lawrence Berkeley National Laboratory, 1 Cyclotron Road, Berkeley, CA 94720, USA}

\begin{abstract}
This work introduces the subensemble acceptance method 2.0~(SAM-2.0) -- a procedure to correct cumulants of a random number distribution inside a subsystem for the effect of exact global conservation of a conserved quantity to which this number is correlated, with applications to  measurements of event-by-event fluctuations in heavy-ion collisions.
The method expresses the corrected cumulants in terms of the cumulants inside and outside the subsystem that are not subject to the exact conservation.
The derivation assumes that all probability distributions associated with the cumulants are peaked at the mean values but are otherwise of arbitrary shape.
The formalism reduces to the original SAM~\cite{Vovchenko:2020tsr} when applied to a coordinate space subvolume of a uniform thermal system.
As the new method is restricted neither to the uniform systems nor to the coordinate space, it is applicable to fluctuations measured in heavy-ion collisions at various collision energies in different momentum space acceptances.
The SAM-2.0 thus brings the experimental measurements and theoretical calculations of event-by-event fluctuations closer together, as the latter are typically performed without the account of exact global conservation.
\end{abstract}

\maketitle


\section{Introduction}
\label{sec:intro}

Event-by-event fluctuations are among the key observables for probing the QCD phase structure experimentally~\cite{Bzdak:2019pkr}, with measurements performed at experiments in a broad collision energy range from HADES-GSI to ALICE-LHC~\cite{Alt:2007jq,Acharya:2019izy,Adam:2019xmk,Adam:2020unf,Adamczewski-Musch:2020slf}.
Proper theoretical modeling of fluctuations in heavy-ion collisions is crucial for correct interpretation of experimental data. 
Exact conservation of quantities like energy-momentum or the conserved QCD charges has a significant influence on many fluctuation observables~\cite{Bleicher:2000ek,Begun:2004gs,Bzdak:2012an,Braun-Munzinger:2016yjz}.
On one hand, the effect of conservation laws prevents meaningful direct comparisons of experimental data to theoretical calculations based on the grand-canonical ensemble, like lattice QCD~\cite{Bazavov:2017dus,Borsanyi:2018grb} or effective theories~\cite{Isserstedt:2019pgx,Fu:2019hdw}.
On the other hand, a proper accounting for exact conservation laws in dynamical models of heavy-ion collisions such as hydrodynamics is challenging, and most of the available implementations are restricted specifically to the model of ideal hadron resonance gas~\cite{Becattini:2003ft,Becattini:2004rq,Schwarz:2017bdg,Oliinychenko:2019zfk,Oliinychenko:2020cmr}~(see, however, a recent Ref.~\cite{Vovchenko:2020kwg} for a Monte Carlo sampling of an interacting hadron resonance gas).

Recently, a subensemble acceptance method~(SAM) has been developed~\cite{Vovchenko:2020tsr,Vovchenko:2020gne} that allows one to correct cumulants of conserved charges for the effect of exact global conservation laws. 
More specifically, the SAM expresses the cumulants measured in a coordinate space subsystem of a thermal system in terms of the grand-canonical susceptibilities and a fraction $\alpha$ of the whole volume covered by the subsystem.
The method has been applied in Ref.~\cite{Vovchenko:2020tsr} to describe baryon number fluctuations at the LHC energies,
where cuts in momentum rapidity can be identified with the subsystem in the longitudinal coordinate space due to Bjorken flow.
The applicability of the SAM to measurements at lower collision energies, like e.g. beam energy scan at RHIC, however, is less clear: the created system is not uniform, with matter properties differing strongly between central and forward-backward rapidity regions~\cite{Li:2018fow} while the absence of the longitudinal boost invariance does not allow one to properly correlate the experimental momentum acceptance with a coordinate space subvolume.

This work introduces the SAM-2.0, which addresses the limitations of the original SAM.
Namely, the method allows one to calculate the effect of exact global conservation on cumulants measured in an arbitrary subsystem, for instance in momentum space as appropriate for the experiment.
Also, the system need not be uniform or equilibrated, the method is applicable in any scenario where all the relevant distributions are highly peaked around the means.
In addition, the SAM-2.0 allows one to evaluate the effect of a global conservation law, such as that of net baryon number, on the cumulants of a non-conserved quantity like the net proton number.
This is particularly relevant for the experiment, where it is usually not possible to measure all particles contributing to a given conserved charge.
A typical use case of the SAM-2.0 would be correcting the  theoretical calculations of the ``grand-canonical'' event-by-event fluctuations within dynamical models of heavy-ion collisions, like hydrodynamics, for global conservation laws.
In the limit of uniform thermal system and fluctuations in the coordinate space subvolume, the SAM-2.0 reduces to the results of the original SAM~\cite{Vovchenko:2020tsr}.

The paper is structured as follows.
The SAM-2.0 is presented in Sec.~\ref{sec:method}, covering fluctuations of both the conserved and non-conserved quantities.
The applications of SAM-2.0 are illustrated in Sec.~\ref{sec:appl} on the example of net proton and net baryon fluctuations at the LHC.
Limitations of the method are discussed in Sec.~\ref{sec:limitations}.
Discussion and summary in Sec.~\ref{sec:summary} close the article.

\section{Method}
\label{sec:method}

Consider fluctuations of a random number, to which one can refer to as a conserved quantity~(charge) $B$, that characterizes the system.
The whole system is partitioned into two parts: the in-system~(the acceptance) and the out-system~(the complement).
The charge $B$ is additive between the subsystems, i.e. the total charge is the sum 
\eq{
B = B_{\rm in} + B_{\rm out}~.
}

First, consider the fluctuations of $B$ in the absence of constraints on the total value of the charge.
The superscript ``gce'' will be used to denote this scenario, in analogy to the grand-canonical ensemble in statistical mechanics.
It should be emphasized, however, that these fluctuations do not necessarily have to come from the grand-canonical statistical mechanics.
Instead, the method is solely based on the following two assumptions regarding the unconstrained distributions:
\begin{enumerate}
    \item The probability distributions $\Pgce_{\rm in}(B_{\rm in})$ and $\Pgce_{\rm out}(B_{\rm out})$ of the charges in the two subsystems are independent, i.e. the joint probability distribution factorizes:
    \eq{\label{eq:indep}
    \Pgce(\Bin,\Bout) = \Pgce_{\rm in}(\Bin) \times \Pgce_{\rm out}(\Bout).
    }
    
    \item $\Pgce_{\rm in}(B_{\rm in})$ and $\Pgce_{\rm out}(B_{\rm out})$ are unimodal distributions that are highly peaked at the mean, such that the maximum term method is applicable. To be more precise, it is assumed that, for any $t$ in the neighborhood of $t = 0$, the following property holds
    \eq{\label{eq:mterm}
    & \ln\left[ \sum_{B_{\inout}} e^{t B_{\inout}} \Pgce_{\inout} (B_{\inout}) \right] = \non
    & \quad \ln\left[ e^{t\,B_{\inout}^{\rm max}(t)} P^{\rm gce}_{\inout} [B_{\inout}^{\rm max}(t)] \right]~.
    }
    Here 
    $B_{\inout}^{\rm max}(t)$
    corresponds to the maximum term of a generalized (unnormalized) probability distribution $\tilde{P}^{\rm gce}_{\inout}(B_{\inout}; t) = e^{t B_{\inout}} \, \Pgce_{\inout}(B_{\inout})$, thus $B_{\inout}^{\rm max}(t = 0)$ corresponds to the maximum of $\Pgce_{\inout}(B_{\inout})$.
\end{enumerate}

\begin{figure}[t]
  \centering
  \includegraphics[width=.49\textwidth]{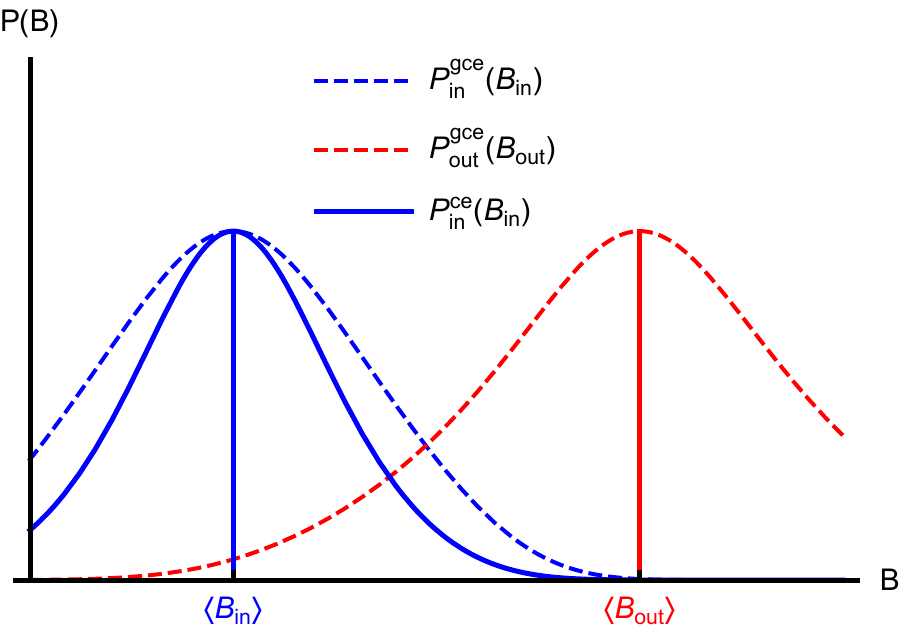}
  \caption{
   The schematic view of the ``grand-canonical''~(dashed lines) and the ``canonical''~(solid line) distributions of the conserved quantity $B$ inside~(blue) and outside~(red) the acceptance.
  }
  \label{fig:schema}
\end{figure}

One can show that the distributions $P^{\rm gce}_{\inout}$ are peaked at the corresponding mean values, i.e. $B_{\inout}^{\rm max}(t=0) = \mean{B_{\inout}}$~(Fig.~\ref{fig:schema}). Indeed,
\eq{\label{eq:meanB}
\mean{B_{\inout}} & = \sum_{B_{\inout}} \, B_{\inout} \, P^{\rm gce}_{\inout} (B_{\inout}) \non
& = \left. \frac{\partial}{\partial t} \ln\left[ \sum_{B_{\inout}} e^{t\,B_{\inout}} \Pgce_{\inout} (B_{\inout})  \right] \right|_{t=0} \non
& = \left. \frac{\partial}{\partial t} \ln\left\{  e^{t\,B_{\inout}^{\rm max}(t)} \Pgce_{\inout} [B_{\inout}^{\rm max}(t)]  \right\} \right|_{t=0} \non
& = B_{\inout}^{\rm max}(t = 0)~,
}
where Eq.~\eqref{eq:mterm} was used, as well as the fact that $d\Pgce_{\inout} (B_{\inout}) / d B_{\inout} = 0$ at $B_{\inout} = B_{\inout}^{\rm max}$.

Without the loss of generality, one can represent the probability distributions as
\eq{\label{eq:Pgcef}
\Pgce_{\inout} (B_{\inout}) = A_{\inout}^{\rm gce} \, e^{-f_{\inout}(B_{\inout})}.
}
Here $A_{\inout}^{\rm gce}$ is the normalization factor while the shape of the whole distribution is fully encoded by  $f_{\inout}(B_{\inout})$.
The fact that the distributions are peaked at the mean can be represented as
\eq{\label{eq:maximum}
\frac{d \Pgce_{\inout}}{d B_{\inout}}(\mean{B_{\inout}}) = 0 \quad \Longleftrightarrow \quad f_{\inout}'(\mean{B_{\inout}})~ = ~ 0,
}
or, more generally, for $t \neq 0$ one can define $\mean{B_{\inout}}(t)$ through $d \tilde{P}^{\rm gce}_{\inout} / d B_{\inout} \, [\mean{B_{\inout}}(t)] = 0$:
\eq{\label{eq:maximum2}  
f_{\inout}'[\mean{B_{\inout}}(t)]~ = ~ t.
}

The relation~\eqref{eq:maximum2} can be used to express cumulants $\kappa_{n,\inout}^{\rm gce}$ of the $\Pgce_{\inout}$ distribution in terms of $f_{\inout}$ and its derivatives. 
This can be achieved by differentiating Eq.~\eqref{eq:maximum2} and observing that, by definition,
\eq{\label{eq:gcedef}
\kappa_{n,\inout}^{\rm gce} \equiv \left. \frac{\partial^{n-1} \mean{B_{\inout}}(t)}{\partial t^{n-1}} \right|_{t=0}.
}
Taking a single derivative with respect to $t$ allows one to calculate $\kappa_{2,\inout}^{\rm gce}$:
\eq{\label{eq:kappa2gce}
\kappa_{2,\inout}^{\rm gce} = \frac{1}{f''_{\inout}(\mean{B_{\inout}})}~.
}
The higher-order cumulants can be obtained by continuously differentiating Eq.~\eqref{eq:kappa2gce} with respect to $t$ where it is implied $\mean{B_{\inout}} \to \mean{B_{\inout}}(t)$.
Appendix~\ref{app:invsingle} presents a recursive procedure to express the derivatives of $f$ in terms of $\kappa_{n,\inout}^{\rm gce}$, which will be useful for the application of SAM.

The exact global conservation of charge $B$ is achieved by rejecting all configurations that do not satisfy this conservation law
\eq{\label{eq:Pcejoint}
\Pce(\Bin, \Bout) \propto \Pgce(\Bin,\Bout) \, \delta_{\Bin+\Bout,B},
}
where $B = \mean{\Bin} + \mean{\Bout}$.
Due to the Kroenecker symbol $\delta_{\Bin+\Bout,B}$, the distributions of $\Bin$ and $\Bout$ that are subject to global conservation are no longer independent of one another.
Here the goal is to evaluate the cumulants $\kappa_{n,\rm in}^{\rm ce}$ of the constrained distribution of $\Bin$ in terms of the cumulants $\kappa_{n,\inout}^{\rm gce}$ of the unconstranined $B_{\inout}$ distributions. The distributions $\Pgce_{\inout}(B_{\inout})$ and $\Pce_{\rm in}(\Bin)$ are depicted schematically in Fig.~\ref{fig:schema}.

A particular physical example which satisfies the assumptions of the method is a statistical mechanics system  with a $U(1)$ conserved charge $B$ in the thermodynamic limit, where the implementation of global conservation laws corresponds to the transition from the grand-canonical to the canonical ensemble characterizing the full system.
In this case the two subsystems may be identified with a partition in the coordinate space whereas the probability functions are proportional to the canonical partition functions, $P \propto Z_{\rm ce} \propto e^{-\frac{V}{T} f}$, where $f$ is the free energy density.
Such a setup forms the basis of the original SAM framework~\cite{Vovchenko:2020tsr}.
Note, however, that the SAM-2.0 presented here is more general and applies to any scenario that satisfies the two assumptions listed above.
In particular, the system does not have to be uniform and in perfect equilibrium whereas the partition into subsystems need not be performed solely in the coordinate space.

\subsection{Cumulants of a conserved quantity}
\label{sec:cons}

The goal is to evaluate the cumulants $\kappa_{n,\rm in}^{\rm ce}$ of the constrained distribution of $\Bin$ in terms of the cumulants $\kappa_{n,\inout}^{\rm gce}$ of the unconstranined $B_{\inout}$ distributions.
Such a mapping shall be denoted by
\eq{\label{eq:SAMmapping}
\{\kappa_{n,\rm in}^{\rm ce}\} = 
\mathcal{S}
\left[\{\kappa_{n,\rm in}^{\rm gce}\}, \{\kappa_{n,\rm out}^{\rm gce}\}\right].
}

It follows from Eq.~\eqref{eq:Pcejoint} that the probability function $\Pce_{\rm in}(B_{\rm in})$ can be written as
\eq{
\Pce_{\rm in}(B_{\rm in}) & = \tilde{A}_{\rm in}^{\rm ce} \sum_{\Bout} \Pgce_{\rm in}(B_{\rm in}) \, \Pgce_{\rm out}(\Bout) \, \delta_{\Bin+\Bout,B} \non
& = \tilde{A}_{\rm in}^{\rm ce} \,  \Pgce_{\rm in}(B_{\rm in}) \, \Pgce_{\rm out}(B - B_{\rm in}) \non
& = A_{\rm in}^{\rm ce} \,   e^{-f_{\rm in}(B_{\rm in})} \, e^{-f_{\rm out}(B - B_{\rm in})}.
}

The distribution $\Pce_{\rm in}(B_{\rm in})$ is peaked at the ``grand-canonical'' mean $\mean{B_{\rm in}}$ given by Eq.~\eqref{eq:meanB}. Indeed, by maximizing $\Pce_{\rm in}(B_{\rm in})$ with respect to $\Bin$ one obtains an equation
\eq{
f_{\rm in}'(B_{\rm in}) = f'_{\rm out}(B - B_{\rm in}),
}
which is satisfied for $B_{\rm in} = \mean{B_{\rm in}}$ as in this case $f_{\rm in}'(\mean{B_{\rm in}}) = 0$ and $f_{\rm out}'(B - \mean{B_{\rm in}}) = f_{\rm out}'(\mean{B_{\rm out}}) = 0$.
Thus, the mean of the $\Bin$ distribution is unaffected by exact global conservation.

To evaluate the 2nd and higher order cumulants one introduces the cumulant generating function
\eq{
G_{\Bin}(t) & = \ln \, \mean{e^{t \Bin}} \non
& = \ln \left \{ \sum_{\Bin} e^{t \Bin} A_{\rm in}^{\rm ce} \, e^{-f_{\rm in}(B_{\rm in})} \, e^{-f_{\rm out}(B - B_{\rm in})} \right\}~.
}
The first order $t$-dependent cumulant $\tilde{\kappa}_{1,\rm in}^{\rm ce}(t) \equiv \mean{B^{\rm ce}_{\rm in}(t)} = \partial G_{\Bin}(t) / \partial t$ corresponds to the maximum of a generalized probability function $\tilde{P}_{\rm in}^{\rm ce} (\Bin;t) = e^{t \Bin} P_{\rm in}^{\rm ce} (\Bin)$.
Maximizing $\tilde{P}_{\rm in}^{\rm ce} (\Bin;t)$ yields the equation
\eq{\label{eq:coreEq}
t = f_{\rm in}'[\mean{B^{\rm ce}_{\rm in}}(t)] - f'_{\rm out}[\mean{B^{\rm ce}_{\rm out}}(t)],
}
where $\mean{B^{\rm ce}_{\rm out}}(t) = B - \mean{B^{\rm ce}_{\rm in}}(t)$.

The higher order cumulants $\kappa_{n,\rm in}^{\rm ce}$ correspond to the derivatives of $\tilde{\kappa}_{1,\rm in}^{\rm ce}(t) = \mean{B_{\inout}^{\rm ce}}(t)$ evaluated at $t = 0$:
\eq{\label{eq:kappanFromLow}
\kappa_{n,\rm in}^{\rm ce} = \left. \frac{\partial^{n-1} \mean{B_{\inout}^{\rm ce}}(t)}{\partial t^{n-1}} \right|_{t=0}.
}
These cumulants can be evaluated recursively, by successively applying the $t$ derivatives to Eq.~\eqref{eq:coreEq} and evaluating them at $t = 0$.
\begin{widetext}
The $n$th order derivative of Eq.~\eqref{eq:coreEq} with respect to $t$ can be obtained by applying Fa\'a di Bruno~(FdB) formula~\cite{comtet2012advanced} to the right-hand-side~(see Appendix~\ref{app:FdB} for details):
\eq{\label{eq:FdB1}
\delta_{n,1} = \sum_{k=1}^n \, f^{(k+1)}_{\rm in} \, B_{n,k} \left( \kappa_{2,\rm in}^{\rm ce}, \ldots, \kappa_{n-k+2,\rm in}^{\rm ce}\right) 
- 
\sum_{k=1}^n \, f^{(k+1)}_{\rm out} \, B_{n,k}\left( -\kappa_{2,\rm in}^{\rm ce}, \ldots, -\kappa_{n-k+2,\rm in}^{\rm ce} \right)~,
}
where Eq.~\eqref{eq:kappanFromLow} was used and $f^{(k+1)}_{\inout} \equiv f^{(k+1)}_{\inout}[\mean{B_{\inout}}]$ is the $(k+1)$th derivative of the function $f_{\inout}$~[Eq.~\eqref{eq:Pgcef}] evaluated at the mean $\mean{B_{\inout}}$.
Here $B_{n,k}$ are Bell polynomials.
One can now separate the term $k = 1$ in Eq.~\eqref{eq:FdB1} from the rest of the sum.
Using the property $B_{n,1}(x_1,\ldots,x_n) = x_n$ one obtains
\eq{\label{eq:FdB2}
\delta_{n,1} = (f''_{\rm in} + f''_{\rm out}) \, \kappa_{n+1,\rm in}^{\rm ce} + \sum_{k=2}^n \left[f^{(k+1)}_{\rm in} \, B_{n,k} \left( \kappa_{2,\rm in}^{\rm ce}, \ldots, \kappa_{n-k+2,\rm in}^{\rm ce}\right) - f^{(k+1)}_{\rm out} \, B_{n,k}\left( -\kappa_{2,\rm in}^{\rm ce}, \ldots, -\kappa_{n-k+2,\rm in}^{\rm ce} \right)\right]~,
}
thus, $\kappa_{n+1,\rm in}^{\rm ce}$ reads
\eq{\label{eq:kappancerec}
\kappa_{n+1,\rm in}^{\rm ce} = \frac{\delta_{n,1} - \sum_{k=2}^n \left[f^{(k+1)}_{\rm in} \, B_{n,k} \left( \kappa_{2,\rm in}^{\rm ce}, \ldots, \kappa_{n-k+2,\rm in}^{\rm ce}\right) - f^{(k+1)}_{\rm out} \, B_{n,k}\left( -\kappa_{2,\rm in}^{\rm ce}, \ldots, -\kappa_{n-k+2,\rm in}^{\rm ce} \right)\right]}{f''_{\rm in} + f''_{\rm out}}~.
}
\end{widetext}

Equation~\eqref{eq:kappancerec} is a recursive relation for $\kappa_{n+1,\rm in}^{\rm ce}$, expressing this quantity in terms of the lower order cumulants up to $\kappa_{n,\rm in}^{\rm ce}$ and the ``grand-canonical'' functions $f_{\inout}$ and their derivatives, which themselves can be expressed in terms of the ``grand-canonical'' cumulants $\kappa_{n,\rm in}^{\rm gce}$~(see Appendix~\ref{app:invsingle}).
This recursive procedure thus defines the mapping function $\mathcal{S}$ in Eq.~\eqref{eq:SAMmapping}.

It is instructive to evaluate the leading cumulants explicitly.
The first-order cumulant corresponds to the mean value $\mean{\Bin}$ which is unaffected by the conservation laws, thus $\kappa_{1,\rm in}^{\rm ce} = \kappa_{1,\rm in}^{\rm gce} = \mean{\Bin}$.
Setting $n = 1$ one obtains the variance
\eq{\label{eq:k2ce}
\kappa_{2,\rm in}^{\rm ce} & = \frac{1}{f''_{\rm in} + f''_{\rm out}} \non
& = \frac{\kappa_{2,\rm in}^{\rm gce} \, \kappa_{2,\rm out}^{\rm gce}}{\kappa_{2,\rm in}^{\rm gce} + \kappa_{2,\rm out}^{\rm gce}},
}
where Eq.~\eqref{eq:kappa2gce} was used for $f''_{\inout}$.
For $n = 2$ one has
\eq{\label{eq:k3ce}
\kappa_{3,\rm in}^{\rm ce} & =
\frac{(\kappa_{2,\rm in}^{\rm ce})^2 \, (f'''_{\rm out} - f'''_{\rm in})}{f''_{\rm in} + f''_{\rm out}} \non
& = (\kappa_{2,\rm in}^{\rm ce})^3 \, 
\left[ 
\frac{\kappa_{3,\rm in}^{\rm gce}}{(\kappa_{2,\rm in}^{\rm gce})^3} - \frac{\kappa_{3,\rm out}^{\rm gce}}{(\kappa_{2,\rm out}^{\rm gce})^3}
\right]~ \non
& = \frac{\kappa_{3,\rm in}^{\rm gce} (\kappa_{2,\rm out}^{\rm gce})^3 - \kappa_{3,\rm out}^{\rm gce} (\kappa_{2,\rm in}^{\rm gce})^3}{\left(\kappa_{2,\rm in}^{\rm gce} + \kappa_{2,\rm out}^{\rm gce}\right)^3}~.
}
Here $f'''_{\inout}$ was taken from Eq.~\eqref{eq:fthreegce}.
The fourth-order cumulant ($n = 3$) is
\eq{
\kappa_{4,\rm in}^{\rm ce} & = 
\frac{1}{\left(\kappa_{2,\rm in}^{\rm gce} + \kappa_{2,\rm out}^{\rm gce}\right)^5} \times \left\{ 
\kappa_{4, \rm in}^{\rm gce} (\kappa_{2, \rm out}^{\rm gce})^5
+ \kappa_{4, \rm out}^{\rm gce} (\kappa_{2, \rm in}^{\rm gce})^5
\right. \non
& \quad + (\kappa_{2,\rm out}^{\rm gce})^4 \, \left[ \kappa_{2,\rm in}^{\rm gce} \kappa_{4,\rm in}^{\rm gce} - 3 (\kappa_{3,\rm in}^{\rm gce})^2 \right]
\non
& \quad + (\kappa_{2,\rm in}^{\rm gce})^4 \, \left[ \kappa_{2,\rm out}^{\rm gce} \kappa_{4,\rm out}^{\rm gce} - 3 (\kappa_{3,\rm out}^{\rm gce})^2 \right]
\non
& \quad - \left. 6 (\kappa_{2,\rm in}^{\rm gce})^2 (\kappa_{2,\rm out}^{\rm gce})^2 \, \kappa_{3, \rm in}^{\rm gce} \, \kappa_{3, \rm out}^{\rm gce} \right\}.
}
Any higher-order cumulant can be obtained from Eq.~\eqref{eq:kappancerec} in the same fashion.
A \texttt{Mathematica} notebook to express cumulant $\kappa_{n,\rm in}^{\rm ce}$ of arbitrary order $n$ in terms of $\{\kappa_{n,\inout}^{\rm gce}\}$ is available via~\cite{SAMgithub}.

It is instructive to consider the following specific case: 
a uniform thermal system of volume $V$ and temperature $T$ which is partitioned into two subsystems of volumes $V_1 = \alpha V$ and $V_2 = (1-\alpha) V$ in the coordinate space.
This was studied in the original SAM paper~\cite{Vovchenko:2020tsr}.
In this case the unconstrained cumulants are determined by the grand-canonical susceptibilities $\chi_n^B$ as follows: $\kappa_{n,\rm in}^{\rm gce} = \alpha V \, T^3 \, \chi_n^B$ and $\kappa_{n,\rm out}^{\rm gce} = (1-\alpha) V \, T^3 \, \chi_n^B$.
Substituting these relations into Eqs.~\eqref{eq:k2ce} and~\eqref{eq:k3ce} one obtains
\eq{
\kappa_{2,\rm in}^{\rm ce} &= V \, T^3 \, \alpha (1-\alpha)  \chi_2^B, \\
\kappa_{3,\rm in}^{\rm ce} &=  V \, T^3 \,\alpha (1-\alpha) (1-2\alpha) \chi_3^B~, \\
\kappa_{4,\rm in}^{\rm ce} &= V \, T^3 \, \alpha \beta \, \left[(1-3\alpha \beta)\chi_4^B - 3 \alpha \beta \frac{(\chi_3^B)^2}{\chi_2^B} \right]~.
}
Here $\beta = 1 - \alpha$.
These relations reproduce the results of Ref.~\cite{Vovchenko:2020tsr}.
It should be emphasized however, that the SAM-2.0 is more general than the original SAM, and allows one to perform the correction for exact conservation for any two distributions that satisfy the assumptions~\eqref{eq:indep} and~\eqref{eq:mterm}.
In particular, the method is applicable to momentum space acceptances.

\subsection{Cumulants of a non-conserved quantity}
\label{sec:noncons}

The global conservation affects not only the cumulants of a quantity which is conserved, but also that of non-conserved numbers that are correlated to the conserved one.
A common example encountered in heavy-ion collisions are cumulants of net proton number.
Even though this quantity is not exactly conserved, the fact that (anti)protons form a subset of all (anti)baryons make the net proton number correlated to the net baryon number and thus being affected by baryon conservation~\cite{Bzdak:2012an}.
This issue is of high practical relevance for heavy-ion collisions because the protons, and not the baryons, can be measured directly in the experiment.
The argument can be extended to other measurements and conserved charges, such as the net kaon number affected by the strangeness conservation.
Fluctuations of non-conserved quantities that are correlated to globally conserved charges have earlier been studied in Ref.~\cite{Vovchenko:2020gne} in uniform thermal systems on the level of second order cumulants.
Here these considerations are extended to arbitrary highly peaked distributions and high-order cumulants.

The non-conserved number will be denoted by $N$.
This number is correlated to a conserved quantity~(charge)~$B$.
In the absence of exact global conservation of $B$, the joint distribution of numbers $N_{\inout}$ and $B_{\inout}$ inside/outside acceptance is characterized by a joint probability function $W^{\rm gce}_{\inout} (B_{\inout},N_{\inout})$.
From the definition it follows that 
\eq{
\Pgce_{\inout}(B_{\inout}) = \sum_{N_{\inout}} W^{\rm gce}_{\inout} (B_{\inout},N_{\inout})~.
}
As before, it is assumed that the joint distributions $W^{\rm gce}_{\rm in} (B_{\rm in},N_{\rm in})$ and $W^{\rm gce}_{\rm out} (B_{\rm out},N_{\rm out})$ are uncorrelated to one another,
\eq{
& W^{\rm gce}(B_{\rm in},N_{\rm in},B_{\rm out},N_{\rm out}) = \non
& \qquad W^{\rm gce}_{\rm in} (B_{\rm in},N_{\rm in}) \, W^{\rm gce}_{\rm out} (B_{\rm out},N_{\rm out}),
}
and that both distributions are highly peaked at the means, such that the maximum term method is applicable.
The distributions $W^{\rm gce}_{\inout} (B_{\inout},N_{\inout})$ are rewritten in the exponential form as follows
\eq{\label{eq:Wgce}
W^{\rm gce}_{\inout} (B_{\inout},N_{\inout}) = C_{\inout}^{\rm gce} \, e^{-g_{\inout}(B_{\inout}, N_{\inout})}.
}

The exact global conservation of $B$ is achieved via the Kronecker delta:
\eq{\label{eq:Wcejoint}
W^{\rm ce}(B_{\rm in},N_{\rm in},B_{\rm out},N_{\rm out}) & \propto W^{\rm gce}(B_{\rm in},N_{\rm in},B_{\rm out},N_{\rm out}) \non
& \quad \times \delta_{\Bin+\Bout,B}.
}
\begin{widetext}
The joint probability function $W^{\rm ce}_{\rm in}(B_{\rm in},N_{\rm in})$ of the numbers $\Bin$ and $N_{\rm in}$ inside the subsystem reads
\eq{
W^{\rm ce}_{\rm in}(B_{\rm in},N_{\rm in}) & = \sum_{\Bout,N_{\rm out}}  W^{\rm gce}_{\rm in} (B_{\rm in},N_{\rm in}) \, W^{\rm gce}_{\rm out} (B_{\rm out},N_{\rm out}) \,  \delta_{\Bin+\Bout,B}
\non
& = W^{\rm gce}_{\rm in} (B_{\rm in},N_{\rm in}) \sum_{N_{\rm out}}  W^{\rm gce}_{\rm out} (B - \Bin,N_{\rm out})
\non
& = W^{\rm gce}_{\rm in} (B_{\rm in},N_{\rm in}) \, \Pgce_{\rm out}(B-B_{\rm in}) \non
& = C_{\rm in}^{\rm gce} \, A_{\rm out}^{\rm gce} \, e^{-g_{\rm in}(B_{\rm in}, N_{\rm in})} \, e^{-f_{\rm out}(B - \Bin)}~.
}
The distribution $W^{\rm ce}_{\rm in}(B_{\rm in},N_{\rm in})$ depends on the ``grand-canonical'' distribution of the conserved charge $B_{\rm out}$ outside of the subsystem but not on the $N_{\rm out}$ distribution of the non-conserved number.

The cumulants of the joint distribution $W^{\rm ce}_{\rm in}(B_{\rm in},N_{\rm in})$ shall be denoted by $\kappa_{n,m;\rm in}^{\rm ce}$.
The goal is to express these cumulants in terms of the joint ``grand-canonical'' cumulants $\kappa_{n,m;\inout}^{\rm gce}$ of the distributions $W^{\rm gce}_{\inout}(B_{\inout},N_{\inout})$, i.e. to find the mapping $\tilde{\mathcal{S}}$ such that
\eq{\label{eq:SAMmappingtwo}
\kappa_{n,m;\rm in}^{\rm ce} = \tilde{\mathcal{S}}\left[\kappa_{n,m;\rm in}^{\rm gce},\kappa_{n,m;\rm out}^{\rm gce} \right]~.
}

The cumulant generating function for the distribution $W^{\rm ce}_{\rm in}(B_{\rm in},N_{\rm in})$ reads
\eq{
G_{\Bin,N_{\rm in}}(t_B,t_N) &= \ln ~ \mean{e^{t_B \Bin + t_N \Nin}} \non
& = \ln \sum_{\Bin,\Nin} e^{t_B \Bin + t_N \Nin} C_{\rm in}^{\rm gce} \, A_{\rm out}^{\rm gce} \, e^{-g_{\rm in}(B_{\rm in}, N_{\rm in})} \, e^{-f_{\rm out}(B - \Bin)}~.
}
Maximizing a generalized probability function
\eq{
\tilde{W}^{\rm ce}_{\rm in}(B_{\rm in},N_{\rm in};t_B,t_N) & = e^{t_B \Bin + t_N \Nin}  W^{\rm ce}_{\rm in}(B_{\rm in},N_{\rm in}) \non
& = e^{t_B \Bin + t_N \Nin} \, C_{\rm in}^{\rm gce} \, A_{\rm out}^{\rm gce} \, e^{-g_{\rm in}(B_{\rm in}, N_{\rm in})} \, e^{-f_{\rm out}(B - \Bin)}
}
with respect to $\Bin$ and $\Nin$ gives the equations defining $\mean{\Bin^{\rm ce}}(t_B,t_N)$ and $\mean{\Nin^{\rm ce}}(t_B,t_N)$:
\eq{\label{eq:tB}
t_B & = g_{\rm in}^{(1,0)} - f'_{\rm out}, \\
\label{eq:tN}
t_N & = g_{\rm in}^{(0,1)},
}
where the following short-hands were introduced 
\eq{
g_{\rm in}^{(n,m)} & = \left. \frac{\partial^{n+m} \left[ g_{\rm in} (\Bin,\Nin) \right]}{\partial (\Bin)^n \partial (\Nin)^m} \right|_{\Bin = \mean{\Bin^{\rm ce}}(t_B,t_N), \, \Nin = \mean{\Nin^{\rm ce}}(t_B,t_N)}~, \\
f^{(n)}_{\rm out} & =  \left. \frac{\partial^n \left[ f_{\rm out} (\Bout) \right]}{\partial (\Bout)^n} \right|_{\Bout = B-\mean{\Bin^{\rm ce}}(t_B,t_N)}~.
}

The cumulants $\kappa_{n,m;\rm in}^{\rm ce}$ can be evaluated via a recursive procedure.
To do that a derivative $\partial^{n+m}/\partial (t_B)^n \partial (t_N)^m$ is applied to Eqs.~\eqref{eq:tB} and \eqref{eq:tN}.
These derivatives can be evaluated with the help of the multivariate Fa\'a di Bruno's formula in the combinatorial form, which presents them as a sum over partitions of a set~(see Appendix~\ref{app:FdB} for the details on the notation).
Applying the FdB formula to Eq.~\eqref{eq:tB} yields
\eq{
\delta_{n,1} \, \delta_{m,0} & = \sum_{\{\pi_1, \pi_2\} \in \Pi_{n+m}^2} g_{\rm in}^{(1+|\pi_1|,|\pi_2|)} \prod_{B \in \pi_1} \kappa_{B_1+1,B_2;\rm in}^{\rm ce} \prod_{B \in \pi_2} \kappa_{B_1,B_2+1;\rm in}^{\rm ce} 
- \sum_{\pi \in \Pi_{n+m}} f_{\rm out}^{(1+|\pi|)} \, (-1)^{|\pi|} \prod_{B \in \pi} \, \kappa_{B_1+1,B_2;\rm in}^{\rm ce}~.
}
The first sum runs over all partitions $\Pi_{n+m}^2$ of a set $\{1,\ldots,n+m\}$ into blocks of two colors while the second sum corresponds to partitions $\Pi_{n+m}$ into blocks of a single color.
One can separate the terms from both sums that correspond to the partitions consisting of single blocks:
\eq{\label{eq:FdBm1}
\delta_{n,1} \, \delta_{m,0} & = [g_{\rm in}^{(2,0)} +f_{\rm out}''] \, \kappa_{n+1,m;\rm in}^{\rm ce} + g_{\rm in}^{(1,1)} \, \kappa_{n,m+1;\rm in}^{\rm ce} \non
& \quad + \sum_{\{\pi_1, \pi_2\} \in \tilde{\Pi}^2_{n+m}} g_{\rm in}^{(1+|\pi_1|,|\pi_2|)} \prod_{B \in \pi_1} \kappa_{B_1+1,B_2;\rm in}^{\rm ce} \prod_{B \in \pi_2} \kappa_{B_1,B_2+1;\rm in}^{\rm ce} - \sum_{\pi \in \tilde{\Pi}_{n+m} } f_{\rm out}^{(1+|\pi|)} \, (-1)^{|\pi|}  \prod_{B \in \pi} \, \kappa_{B_1+1,B_2;\rm in}^{\rm ce}~.
}
Here $\tilde{\Pi}_{n+m} = \Pi_{n+m} \setminus \{|\pi| = 1\}$ and  $\tilde{\Pi}_{n+m}^2 = \Pi_{n+m}^2 \setminus \{|\pi_1| + |\pi_2| = 1\}$.
Note that the sums in the r.h.s of Eq.~\eqref{eq:FdBm1} contain cumulants $\kappa^{\rm ce}_{\rm in}$ only up to order $n+m$.

Applying the derivative $\partial^{n+m}/\partial (t_B)^n \partial (t_N)^m$ to Eq.~\eqref{eq:tN} and performing the same manipulations one obtains
\eq{\label{eq:FdBm2}
\delta_{n,0} \, \delta_{m,1} & =  g_{\rm in}^{(1,1)} \, \kappa_{n+1,m;\rm in}^{\rm ce} + g_{\rm in}^{(0,2)} \, \kappa_{n,m+1;\rm in}^{\rm ce}  + \sum_{\{\pi_1, \pi_2\} \in \tilde{\Pi}^2_{n+m}} g_{\rm in}^{(|\pi_1|,1+|\pi_2|)} \prod_{B \in \pi_1} \kappa_{B_1+1,B_2;\rm in}^{\rm ce} \prod_{B \in \pi_2} \kappa_{B_1,B_2+1;\rm in}^{\rm ce}~.
}
Equations~\eqref{eq:FdBm1} and~\eqref{eq:FdBm2} can be cast in the form of a system of linear equations for $\kappa_{n+1,m;\rm in}^{\rm ce}$ and $\kappa_{n,m+1;\rm in}^{\rm ce}$:
\eq{\label{eq:syst1}
[g_{\rm in}^{(2,0)} +f_{\rm out}''] \, \kappa_{n+1,m;\rm in}^{\rm ce} + g_{\rm in}^{(1,1)} \, \kappa_{n,m+1;\rm in}^{\rm ce} & = \Sigma_{n,m}^1, \\
\label{eq:syst2}
g_{\rm in}^{(1,1)} \, \kappa_{n+1,m;\rm in}^{\rm ce} + g_{\rm in}^{(0,2)} \, \kappa_{n,m+1;\rm in}^{\rm ce} & = \Sigma_{n,m}^2,
}
where
\eq{\label{eq:lin1}
\Sigma_{n,m}^1 & = \delta_{n,1} \, \delta_{m,0} - \sum_{\{\pi_1, \pi_2\} \in \tilde{\Pi}^2_{n+m}} g_{\rm in}^{(1+|\pi_1|,|\pi_2|)} \prod_{B \in \pi_1} \kappa_{B_1+1,B_2;\rm in}^{\rm ce} \prod_{B \in \pi_2} \kappa_{B_1,B_2+1;\rm in}^{\rm ce} \non
& \quad + \sum_{\pi \in \tilde{\Pi}_{n+m} } f_{\rm out}^{(1+|\pi|)} \, (-1)^{|\pi|}  \prod_{B \in \pi} \, \kappa_{B_1+1,B_2;\rm in}^{\rm ce}, \\
\label{eq:lin2}
\Sigma_{n,m}^2 & = \delta_{n,0} \, \delta_{m,1}  - \sum_{\{\pi_1, \pi_2\} \in \tilde{\Pi}^2_{n+m}} g_{\rm in}^{(|\pi_1|,1+|\pi_2|)} \prod_{B \in \pi_1} \kappa_{B_1+1,B_2;\rm in}^{\rm ce} \prod_{B \in \pi_2} \kappa_{B_1,B_2+1;\rm in}^{\rm ce}~.
}
Solving the system of equations allows one to express the ``canonical'' cumulants $\kappa_{n+1,m;\rm in}^{\rm ce}$ and $\kappa_{n,m+1;\rm in}^{\rm ce}$ in terms of lower-order cumulants as well the derivatives of the ``grand-canonical'' distribution functions $g_{\rm in}(\Bin,\Nin)$ and $f_{\rm out}(\Bout)$.
The derivatives of $f_{\rm out}$ and $g_{\rm in}$ can be expressed in terms of the ``grand-canonical'' cumulants $\kappa^{\rm gce}_{\inout}$~(see Appendices \ref{app:invsingle} \& \ref{app:invtwo}, respectively), i.e. this defines the mapping function $\tilde{\mathcal{S}}$ in Eq.~\eqref{eq:SAMmappingtwo}.
This completes the recursive procedure to evaluate the joint cumulants $\kappa^{\rm ce}_{\rm in}$ of the $(\Bin,\Nin)$ distribution constrained by exact global conservation of charge $B$.

\end{widetext}

The leading order cumulants can be written down as follows. The first order cumulants -- the means -- are unaffected by the exact global conservations, thus
\eq{
\kappa_{1,0;\rm in}^{\rm ce} & = \kappa_{1,0;\rm in}^{\rm gce} = \mean{\Bin}, \\
\kappa_{0,1;\rm in}^{\rm ce} & = \kappa_{0,1;\rm in}^{\rm gce} = \mean{\Nin}.
}
For the second order cumulants~[$n+m = 1$ in Eqs.~\eqref{eq:syst1},~\eqref{eq:syst2}] one has
\eq{\label{eq:kappa2Bmult}
\kappa_{2,0;\rm in}^{\rm ce} & = \frac{\kappa_{2,0;\rm in}^{\rm gce} \, \kappa_{2,0;\rm out}^{\rm gce}}{\kappa_{2,0;\rm in}^{\rm gce} + \kappa_{2,0;\rm out}^{\rm gce}}, \\
\kappa_{1,1;\rm in}^{\rm ce} & = \frac{\kappa_{1,1;\rm in}^{\rm gce} \, \kappa_{2,0;\rm out}^{\rm gce}}{\kappa_{2,0;\rm in}^{\rm gce} + \kappa_{2,0;\rm out}^{\rm gce}},\\
\kappa_{0,2;\rm in}^{\rm ce} & = \kappa_{0,2;\rm in}^{\rm gce} - \frac{(\kappa_{1,1;\rm in}^{\rm gce})^2}{\kappa_{2,0;\rm in}^{\rm gce} + \kappa_{2,0;\rm out}^{\rm gce}}~.
}
Note that Eq.~\eqref{eq:kappa2Bmult} for the variance of the conserved quantity $B$ coincides with the corresponding result~\eqref{eq:k2ce} of Sec.~\ref{sec:cons}, as it should.
A \texttt{Mathematica} notebook to calculate the explicit expressions for the higher-order cumulants $\kappa_{n,m;\rm in}^{\rm ce}$ is available via~\cite{SAMgithub}.

\section{Application: Cumulants of net proton and net baryon distributions in heavy-ion collisions at the LHC}
\label{sec:appl}

To illustrate the formalism it is applied here to correct the cumulants of the net-proton and net-baryon number in different rapidity windows in heavy-ion collisions at the LHC for baryon number conservation.
More specifically Pb-Pb collisions at $\sqrt{s_{\rm NN}} = 2.76$~TeV are considered here.
In Ref.~\cite{Vovchenko:2020kwg} the fluctuations were analyzed using Monte Carlo sampling of hadrons at a blast-wave particlization hypersurface, with account for effects of baryon excluded volume, thermal smearing, and exact global conservation of baryon number.
Here the fluctuations are first calculated analytically in the grand-canonical limit, and then the SAM-2.0 is applied to perform the correction for baryon conservation. 
Comparisons with the Monte Carlo data of Ref.~\cite{Vovchenko:2020kwg} are then performed to verify the validity of the method.

The fluctuations of net proton and net baryon number studied here correspond to measurements around midrapidity as a function of rapidity cut $|y| < \Delta Y_{\rm acc} / 2$.
To account for the thermal smearing, it is assumed that the longitudinal rapidity $y$ of each baryon is smeared at particlization around its space-time rapidity coordinate $\eta_s$ by a Gaussian.
Under this assumption it is possible to calculate analytically the cumulants $\kappa_{n,\rm in}^{B^{\pm}, \rm gce}(\Delta Y_{\rm acc})$ of baryons and antibaryons in the given rapidity acceptance $\Delta Y_{\rm acc}$ in the grand-canonical limit.
The cumulants $\kappa_{n,\rm out}^{B^{\pm}, \rm gce}(\Delta Y_{\rm acc})$ of (anti)baryons outside the acceptance $\Delta Y_{\rm acc}$ are evaluated in the same fashion.
This procedure is described in detail in the Appendix of Ref.~\cite{Vovchenko:2020kwg}~[see Eq.~(93) there] and the results of that paper are used here.

In order to apply the SAM-2.0 to correct net proton and net baryon cumulants for global baryon conservation one first needs to evaluate the corresponding grand-canonical joint cumulants $\kappa_{n,m;\rm in}^{B,p, \rm gce}(\Delta Y_{\rm acc})$ and $\kappa_{n,m;\rm out}^{B,p, \rm gce}(\Delta Y_{\rm acc})$ inside and outside the acceptance $\Delta Y_{\rm acc}$.
The procedure to evaluate $\kappa_{n,m;\inout}^{B,p, \rm gce}(\Delta Y_{\rm acc})$ is the same both for inside and outside the acceptance, thus the focus here is on the former case.

Let $P(N_B)$ be the probability to have $N_B$ baryons in the acceptance. 
This distribution is characterized by the cumulants $\kappa_{n,\rm in}^{B^+, \rm gce}(\Delta Y_{\rm acc})$.
Each baryon ends up as a proton in the final state with a probability~$q \approx 0.33$~\cite{Vovchenko:2020kwg}.
As the model does not contain isospin-dependent correlations at particlization, the selection of final state protons from all baryons is described by the binomial distribution~\cite{Kitazawa:2011wh,Kitazawa:2012at}:
\eq{
P(N_B,N_p) = B(N_B,N_p;q) \, P(N_B)~,
}
where
\eq{
B(N_B,N_p;q) = \binom{N_B}{N_p} \, q^{N_p} (1-q)^{N_B-N_p}~.
}

The cumulant generating function for the joint $(N_B$,$N_p)$ distribution reads
\eq{\label{eq:GtBtp}
G(t_B,t_p) & = \ln \mean{e^{t_B N_B} e^{t_p N_p}} \non
& = \sum_{N_B} P(N_B) e^{t_B N_B} \sum_{N_p} \, B(N_B,N_p;q) \, e^{t_p N_p} \non
& = \sum_{N_B} P(N_B) e^{\gamma(t_B,t_p,q) \, N_B} \non
& = G_B[\gamma(t_B,t_p,q)]~.
}
Here $G_B$ is the cumulant generating function of $N_B$ distribution and
\eq{\label{eq:gammbino}
\gamma(t_B,t_p,q) = t_B + \ln[1-(1-e^{t_p})q]~.
}
To obtain Eq.~\eqref{eq:gammbino} an identity $\sum_{N_p = 0}^{N_B} B(N_B,N_p;q) \, e^{t_p N_p} = [1 - (1 - e^{t_p}) \, q]^{N_B}$ was used.

The joint cumulants $\kappa_{n,m;\inout}^{B^+,p^+, \rm gce}(\Delta Y_{\rm acc})$ of baryon-proton distribution correspond to the Taylor expansion coefficients of the generating function $G(t_B,t_p)$ around $t_B = t_p = 0$.
These are evaluated with the help of Fa\`a di Bruno's formula applied to Eq.~\eqref{eq:GtBtp} and expressed in terms of the cumulants $\kappa_{n,\rm in}^{B^+, \rm gce}(\Delta Y_{\rm acc})$ of the $N_B$ distribution as follows:
\eq{\label{eq:kappapBtherm}
& \kappa_{n,m;\rm in}^{B^+,p^+, \rm gce}(\Delta Y_{\rm acc})  = \delta_{m,0} \, \kappa_{n,\rm in}^{B^+, \rm gce}(\Delta Y_{\rm acc}) \non
& \quad + \sum_{k=1}^m \, \kappa_{n+k,\rm in}^{B^+, \rm gce}(\Delta Y_{\rm acc}) \, B_{n,k}(\gamma'_{t_p}, \ldots, \gamma^{(m-k+1)}_{t_p})~.
}

The same procedure applies for the joint cumulants $\kappa_{n,m;\rm in}^{B^-,p^-, \rm gce}(\Delta Y_{\rm acc})$ of the antiproton-antibaryon distribution in the acceptance.

To obtain the joint cumulants $\kappa_{n,m;\inout}^{B,p, \rm gce}(\Delta Y_{\rm acc})$ of the accepted net-proton/net-baryon distribution one makes use of the following property of the model: there are no grand-canonical correlations between baryons and antibaryons.
In this case one obtains
\eq{
\kappa_{n,m;\inout}^{B,p, \rm gce}(\Delta Y_{\rm acc}) & = \kappa_{n,m;\inout}^{B^+,p^+, \rm gce}(\Delta Y_{\rm acc}) \non
& \quad + (-1)^{n+m} \, \kappa_{n,m;\inout}^{B^-,p^-, \rm gce}(\Delta Y_{\rm acc})~.
}
Note that in cases where the grand-canonical correlations between baryons and antibaryons are present, the procedure to evaluate $\kappa_{n,m;\inout}^{B,p, \rm gce}(\Delta Y_{\rm acc})$ will be more involved and will require the use factorial moments of baryon and antibaryon distributions~\cite{Bzdak:2012ab}.

After evaluating $\kappa_{n,m;\inout}^{B,p, \rm gce}(\Delta Y_{\rm acc})$, these cumulants are inserted into the recursive procedure defined by Eqs.~\eqref{eq:syst1} and~\eqref{eq:syst2} to evaluate the cumulants $\kappa_{n,m;\rm in}^{B,p, \rm ce}(\Delta Y_{\rm acc})$ of accepted net-proton/net-baryon distribution that are constrained by exact baryon conservation.
The following three ratios of cumulants are analyzed, for both the net protons,
\eq{
& \frac{\kappa_{2,\rm in}^{p,\rm ce}}{\mean{p^+} + \mean{p^-}}, \qquad \frac{\kappa_{4,\rm in}^{p,\rm ce}}{\kappa_{2,\rm in}^{p,\rm ce}}, \qquad \frac{\kappa_{6,\rm in}^{p,\rm ce}}{\kappa_{2,\rm in}^{p,\rm ce}} ,
}
and net baryons
\eq{
& \frac{\kappa_{2,\rm in}^{B,\rm ce}}{\mean{B^+} + \mean{B^-}}, \qquad \frac{\kappa_{4,\rm in}^{B,\rm ce}}{\kappa_{2,\rm in}^{B,\rm ce}}, \qquad \frac{\kappa_{6,\rm in}^{B,\rm ce}}{\kappa_{2,\rm in}^{B,\rm ce}}~,
}
as a function of rapidity acceptance $\Delta Y_{\rm acc}$.
Here 
\eq{
\kappa_{n,\rm in}^{p,\rm ce} \equiv \kappa_{0,n;\rm in}^{B,p, \rm ce} \quad \text{and} \quad \kappa_{n,\rm in}^{B,\rm ce} \equiv \kappa_{n,0;\rm in}^{B,p, \rm ce}~.
}

\begin{figure}[t]
  \centering
  \includegraphics[width=.49\textwidth]{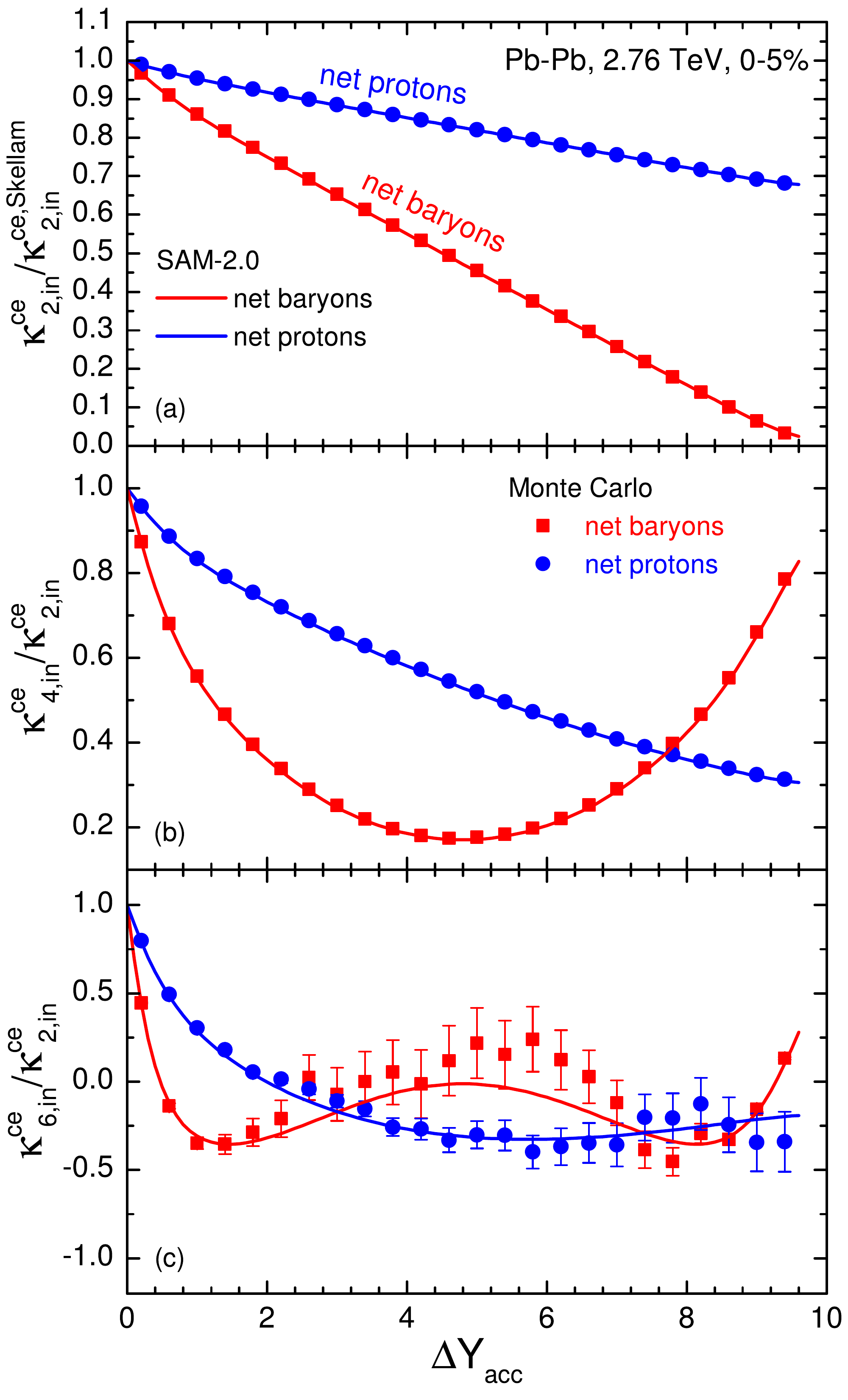}
  \caption{
   Rapidity acceptance dependence of net baryon~(red squares) and net proton~(blue circles) cumulant ratios (a) $\kappa_{2,\rm in}^{\rm ce}/\kappa_{2,\rm in}^{\rm ce,Skellam}$, (b) $\kappa_{4,\rm in}^{\rm ce}/\kappa_{2,\rm in}^{\rm ce}$, and (c) $\kappa_{6,\rm in}^{\rm ce}/\kappa_{2,\rm in}^{\rm ce}$ in 0-5\% central Pb-Pb collisions at the LHC evaluated using within the model of Ref.~\cite{Vovchenko:2020kwg}.
   The symbols depict the Monte Carlo results of Ref.~\cite{Vovchenko:2020kwg}.
   The solid lines  correspond to the analytical calculations where the correction for baryon conservation has been applied within the SAM-2.0 framework.
  }
  \label{fig:netbaryonproton}
\end{figure}

Figure~\ref{fig:netbaryonproton} depicts the results of the calculations.
The analytical SAM-2.0 results are compared to the Monte Carlo results of Ref.~\cite{Vovchenko:2020kwg}, shown by the symbols.
It is seen that the SAM-2.0 results are consistent with the Monte Carlo sampling procedure, for all values of $\Delta Y_{\rm acc}$.
This validates the method for conditions realized at the LHC energies and thus allows one to forgo the computationally expensive Monte Carlo sampling for calculating the high-order cumulants of (net-)proton number distribution.

The results also illustrate some of the advantages of the SAM-2.0 over the original SAM.
In particular, the SAM-2.0 made it possible to compute analytically the net proton cumulants, which cannot be done using the original SAM since net proton is not a conserved quantity.
For the net baryons the advantages are not as evident at the LHC, as the created system is largely uniform due to the longitudinal boost invariance, making also the original SAM applicable for net baryons. 
The differences between SAM-2.0 and SAM become more important at lower collision energies where the created fireball is no longer uniform.
Applications of the SAM-2.0 framework for Au-Au collisions at RHIC beam energy scan can be found in Ref.~\cite{BESflucs}.

\section{Limitations}
\label{sec:limitations}

The SAM-2.0 is based on the two assumptions introduced at the beginning of Sec.~\ref{sec:method}, namely that the unconstrained conserved charge distributions inside and outside the acceptance are independent of one another~[Eq.~\eqref{eq:indep}], and that these distributions are highly peaked at the mean values, such that the maximum term method is applicable~[Eq.~\eqref{eq:mterm}].
In practice, the conditions \eqref{eq:indep} and \eqref{eq:mterm} may not always be satisfied fully, and if that is the case, the true results may deviate from the SAM-2.0 expressions.
These possible deviations are discussed in this section.

\subsection{The peakedness of the distributions}

The relation~\eqref{eq:mterm} is accurate in large statistical systems and it becomes exact in the infinite volume limit.
In particular, this is the case when the total conserved charge $B$ is large compared to unity.
Note that the method is also accurate when $B$ is small~(or even vanishing like the net strangeness in heavy-ion collisions), but the system volume is sufficiently large.
To see why this is the case one can consider a system of particles and antiparticles with vanishing net conserved charge, $B = N_B - N_{\bar{B}} = 0$~\cite{Begun:2004gs}.
Even though in this case the net charge is zero, the numbers of particles and antiparticles are large, $N_B,\,N_{\bar{B}} \gg 1$, and the $B$ distribution is highly peaked around this small value of $B$.

As far as heavy-ion applications are concerned, the finite volume effects are essentially negligible in central collisions of heavy ions across a broad energy range~\cite{Bzdak:2012an,Vovchenko:2020kwg}.
The only possible exceptions to this are the vicinity of a critical point, where the multiplicity distributions become broad, or a first-order phase transition, where the distributions may be bimodal. 
In such cases, as well as in the case of small systems or net strangeness at low energies~\cite{Hamieh:2000tk}, the accuracy of the relation~\eqref{eq:mterm} should be verified explicitly.

\subsection{The independence of the in/out distributions}

The assumption~~\eqref{eq:indep} of independence of $\Pgce_{\rm in}(\Bin)$ and $\Pgce_{\rm out}(\Bin)$ is more subtle. 
As follows from Fig.~\ref{fig:netbaryonproton}, this assumption appears to hold true for baryon number distributions in rapidity acceptances at the LHC energies, where there is strong correlation the longitudinal momenta and coordinates due to Bjorken flow.
One can now consider the extreme opposite: No correlations between momenta and coordinates of particles.
To be more specific, consider a static thermal system of particles where $B$ plays the role of the conserved particle number in the system.\footnote{Compared to the baryon number in heavy-ion collisions, here the production of antiparticles is not considered. Such as a setup thus resembles nucleon number fluctuations in heavy-ion collisions at moderate energies, where antibaryons can be neglected.}
It is assumed that correlations between particles may only depend on their coordinates, but not on their momenta.
As shown below, this represents an example of the worst-case scenario for the accuracy of SAM-2.0.

Consider the fluctuations of particle number in a particular momentum acceptance $\Delta p$.
As the absence of momentum-dependent correlations between particles has been assumed, the probability $\alpha$ that each particle ends up in the acceptance $\Delta p$ is independent of all other particles and corresponds to the ratio of mean number of accepted particles $\mean{\Bin}$ to the total number $B$:
\eq{
\alpha = \frac{\mean{\Bin}}{B}~.
}
The $\Bin$ cumulants are then given by the binomial distribution and the first three cumulants read:
\eq{\label{eq:binoce1}
\kappa_{1,\rm in}^{\rm ce} & = \alpha B, \\
\label{eq:binoce2}
\kappa_{2,\rm in}^{\rm ce} & = \alpha (1-\alpha) \, B, \\
\label{eq:binoce3}
\kappa_{3,\rm in}^{\rm ce} & = \alpha (1-\alpha) (1-2\alpha) \, B.
}
These expressions are independent of the equation of state, i.e. they are independent of the grand-canonical cumulants $\kappa_{n,\inout}^{\rm gce}$ that would characterize the system in the absence of exact global conservation.

One can now calculate these cumulants using the SAM-2.0.
In the absence of global conservation, the total particle number $B$ fluctuates and this is characterized by the grand-canonical cumulants $\kappa_{n}^{\rm gce}$.
As the probability that a given particle ends up inside or outside the momentum space acceptance $\Delta p$ is independent from all other particles,
the grand-canonical cumulants $\kappa_{n,\inout}^{\rm gce}$ 
are obtained by convoluting the underlying distributions described by $\kappa_{n}^{\rm gce}$ with the binomial distribution~\cite{Kitazawa:2012at,Bzdak:2012ab,Savchuk:2019xfg}.
For example, the leading three $\kappa_{n,\rm in}^{\rm gce}$ read
\eq{\label{eq:bino1}
\kappa_{1,\rm in}^{\rm gce} & = \alpha \, \kappa_{1}^{\rm gce}, \\
\label{eq:bino2}
\kappa_{2,\rm in}^{\rm gce} & = \alpha^2 \, \kappa_2^{\rm gce} + \alpha (1-\alpha) \, \kappa_{1}^{\rm gce}, \\
\label{eq:bino3}
\kappa_{3,\rm in}^{\rm gce} & = \alpha^3 \, \kappa_3^{\rm gce} + \alpha (1-\alpha) \, \left[3 \, \alpha \kappa_2^{\rm gce} + (1-2\alpha) \kappa_{1}^{\rm gce} \right].
}
The expressions for $\kappa_{n,\rm out}^{\rm gce}$ are obtained by a substitution $\alpha \to 1 - \alpha$.

If the distributions of $\Bin$ and $\Bout$ were independent, the sum of the $\inout$ cumulants would give the total cumulant.
One obtains instead
\eq{
\kappa_{1,\rm in}^{\rm gce} + \kappa_{1,\rm out}^{\rm gce} & = \kappa_{1}^{\rm gce}, \\
\kappa_{2,\rm in}^{\rm gce} + \kappa_{2,\rm out}^{\rm gce} & = \kappa_{2}^{\rm gce} (1-2\alpha\beta)+ 2 \, \alpha \beta \, \kappa_{1}^{\rm gce} \neq \kappa_{2}^{\rm gce}, \\
\kappa_{3,\rm in}^{\rm gce} + \kappa_{3,\rm out}^{\rm gce} & = \kappa_{3}^{\rm gce} (1-3\alpha\beta) + 3 \, \alpha \beta \,  \kappa_{2}^{\rm gce} \neq \kappa_{3}^{\rm gce}.
}
Here $\beta = 1 - \alpha$.
The $\inout$ cumulants are not additive for $n \geq 2$ in the general case, thus the distributions of $\Bin$ and $\Bout$ are not independent.
The only partial case where the cumulants are additive is when the distribution of particle number $B$ follows the Poisson distribution, i.e. $\kappa_n^{\rm gce} = B$ for all $n \geq 1$.

Because the assumption of the independence of the $\Bin$ and $\Bout$ distributions does not hold in the presence of grand-canonical correlations between particles, the SAM-2.0 will not yield the expected exact result for the canonical cumulants of $\Bin$ given by Eqs.~\eqref{eq:binoce1}-\eqref{eq:binoce3}.
The difference between the SAM-2.0 and the exact result for the second order cumulant can be evaluated.
One has $\kappa_1^{\rm gce} = B$ and $\kappa_2^{\rm gce} = B + \delta$, where $\delta \neq 0$ corresponds to the presence of correlations between particles.
The difference between the SAM-2.0 result, evaluated using~\eqref{eq:k2ce}, and the exact result~\eqref{eq:binoce2} is
\eq{\label{eq:compexact}
\left(\kappa_{2,\rm in}^{\rm ce}\right)_{\rm SAM-2.0} - \left(\kappa_{2,\rm in}^{\rm ce}\right)_{\rm exact} = \frac{\alpha^2 \beta^2 \delta (2B + \delta)}{B + \delta - 2 \alpha \beta \delta}~.
}
It follows from Eq.~\eqref{eq:compexact} that, for $|\delta| < B$, the sign of  $\left(\kappa_{2,\rm in}^{\rm ce}\right)_{\rm SAM-2.0} - \left(\kappa_{2,\rm in}^{\rm ce}\right)_{\rm exact}$ is determined by the sign of $\delta$,
thus the SAM-2.0 overestimates the effect of  correlations~(which in this example should not contribute at all) to $\kappa_{2,\rm in}^{\rm ce}$. 
Therefore, when the independence of the distributions $\Pgce_{\rm in}(\Bin)$ and $\Pgce_{\rm out}(\Bin)$ does not hold, the results of the SAM-2.0 may be taken as an estimate for the maximum possible effect of grand-canonical correlations in the cumulants of accepted particles constrained by global conservation.
Note that the largest deviations occur $\alpha = 1/2$ and disappear as $\alpha \to 0$ or $\alpha \to 1$.

In the context of heavy-ion applications, it should be noted that the correlation between the momenta and coordinates of particles due to Bjorken flow diminishes as collision energy is decreased.
The applications of SAM-2.0 at moderate collision energies, for instance below those of RHIC-BES~($\sNN < 7.7$~GeV), should thus be performed with care.

\section{Discussion and summary}
\label{sec:summary}

This work introduced a generalized subensemble acceptance method -- the SAM-2.0 -- to correct cumulants of both conserved and non-conserved quantities in a subsystem~(acceptance) for exact global conservation of a single conserved quantity.
The method expresses the corrected cumulants in terms of the uncorrected cumulants inside and outside the acceptance.
In contrast to the original SAM~\cite{Vovchenko:2020tsr}, the new method is applicable to non-uniform systems and momentum space acceptances, making it suitable for applications in heavy-ion collisions at various collision energies.

The practical use of the SAM-2.0 is in applying a correction for global charge conservation to theoretical calculations of various event-by-event fluctuations.
One can take hydrodynamical simulations of heavy-ion collisions as an example:
\begin{enumerate}
    \item First one calculates the cumulants of interest, both inside and outside the acceptance, without the effect of global conservation. 
    In this case the standard grand-canonical particlization is performed, where emission of particles from every hypersurface element is calculated independently.
    \item Then, one applies the SAM-2.0 to correct the cumulants for exact global conservation.
\end{enumerate}
In such a way one calculates the cumulants affected by global conservation laws without the need to apply a time-consuming, if not intractable, procedure to account for exact global conservation laws at particlization.
The method requires that the unconstrained probability distributions inside and outside the acceptance are independent~[Eq.~\eqref{eq:indep}] and highly peaked~[Eq.~\eqref{eq:mterm}], but have otherwise arbitrary properties.

It should be noted that only the total baryon number (as well as electric charge) is conserved exactly in heavy-ion collisions. This number contains contributions from participants and spectators, which both fluctuate.
In the context of hydrodynamic simulations, the correction for baryon conservation should thus be applied to calculations for a fixed number of participants and then folded with participant fluctuations, as discussed in Ref.~\cite{Bzdak:2016jxo}.
Note that the effect of participant fluctuations is less relevant for central collisions, where the number of spectators is small, as well as in those cases where the experimental data have been corrected for participant fluctuations.

It has been shown here that the SAM-2.0 accurately reproduces the effect of global baryon conservation on cumulants of net proton number fluctuations in Pb-Pb collisions at the LHC.
This has been achieved by comparing the SAM-2.0 results with the direct Monte Carlo sampling.
Applications of the SAM-2.0 to proton number cumulants in Au-Au collisions at RHIC-BES can be found in Ref.~\cite{BESflucs}.

Here only a single globally conserved quantity was considered.
This is sufficient in cases where a single conservation law has a dominant effect on the observable of interest.
The method can be extended to multiple conserved quantities by utilizing cumulants of their joint distributions, similarly as it was done in Ref.~\cite{Vovchenko:2020gne} for the original SAM.


\begin{acknowledgments}

The author thanks V.~Koch for fruitful discussions and useful comments and R.V.~Poberezhnyuk, O.~Savchuk, and C.~Shen for collaboration on related projects.
This work was supported through the
Feodor Lynen Program of the Alexander von Humboldt
foundation, the U.S. Department of Energy, 
Office of Science, Office of Nuclear Physics, under contract number 
DE-AC02-05CH11231231, and within the framework of the
Beam Energy Scan Theory (BEST) Topical Collaboration.

\end{acknowledgments}

\appendix

\section{Partitions of a set and  Fa\`a di Bruno's formula}
\label{app:FdB}

Fa\`a di Bruno's~(FdB) formula is an identity generalizing the chain rule to higher derivatives.
In case of multivariate derivatives, the ``combinatorial'' form of the FdB formula is particularly useful, which involves summation over partitions of a set.

\subsection{Partitions of a set into uncolored blocks}

A partition of a set is a grouping of its elements into non-empty subsets -- the blocks.
One can denote $\Pi_n$ as the set of all partitions of a set $\{1,\ldots,n\}$~\cite{rota1964number}.
For example, the set $\{ 1 \}$ has only a single partition, thus $\Pi_1 = \{ \{ \{1\} \} \}$.
On the other hand, there are two different partitions possible for the set $\{1,2\}$, either both elements are in the same block, or they are split into two different blocks:
\eq{
\Pi_{2} = \{ \{ \{1,2\} \}, \{ \{1\}, \{2\} \} \}.
}
All partitions of a set $\Pi_{n+1}$ can be constructed recursively, by utilizing the previously calculated partitions  $\Pi_{n}$.
To construct $\Pi_{n+1}$ one iterates over all partitions $\pi$ in $\Pi_{n}$ and either adds the element $n+1$ to one of the blocks of $\pi$ or adds a new block to $\pi$ which contains the element $n+1$. 
In this way one iterates over all valid partitions $\Pi_{n+1}$ of a set $\{1,\ldots,n+1\}$.

The partitions of a set $\{1,\ldots,n\}$ can be used to evaluate the following $n$th order derivative of a composite function $f[g(x)]$:
\eq{\label{eq:FdBcomb}
\frac{d^n f[g(x)]}{d x^n} = \sum_{\pi \in \Pi_n} f^{(|\pi|)}[g(x)] \, \prod_{B \in \pi} \, g^{(|B|)}(x)~.
}
Here $\pi$ runs over all partititions $\Pi_n$ of a set $\{1,\ldots,n\}$, $|\pi|$ is the number of blocks in the partition $\pi$, $B$ runs over all blocks in the partition $\pi$, and $|B|$ is the number of elements in the block $B$.
Equation~\eqref{eq:FdBcomb} can be derived by successive applications of the chain, product, and sum rules, that are necessary to compute the higher derivatives.
This expression is equivalent to the more common form of the FdB formula in terms of the Bell polynomials:
\eq{\label{eq:FdBstand}
\frac{d^n f[g(x)]}{d x^n} = \sum_{k=1}^n \, f^{(k+1)} [g(x)] \, B_{n,k} \left[g'(x), \ldots, g^{(n-k+1)}(x)\right] ~.
}

\subsection{Partitions of a set into colored blocks}

Partitions of a set into \emph{colored}~(or \emph{labeled}) blocks generalize the standard partitions $\Pi_n$ into a single type of blocks~\cite{haiman1989incidence}.
This extension allows one to generalize the FdB formula to derivatives of multivariate composite functions.
Assume that each block can be labeled by one of $l$ distinct colors and denote $\Pi^l_n$ as the set of all partitions of a set $\{1,\ldots,n\}$ into such colored blocks.
E.g. for $l = 2$ one has
\eq{
\Pi^2_1 & = \{ \{ \{1\}_1 \}, \{ \{1\}_2 \} \}, \\
\Pi^2_2 & = \{ \{ \{1,2\}_1 \}, \{ \{1,2\}_2 \},\{ \{1\}_1, \{2\}_1 \}, \{ \{1\}_2, \{2\}_2 \}, \non
& \qquad  \{ \{1\}_1, \{2\}_2 \}, \{ \{1\}_2, \{2\}_1 \} \}.
}
Here the subscripts in the right-hand-sides encode the label~($l = 1$ or $2$) of each block.
Similarly to the partitions into unlabeled blocks, $\Pi^l_{n+1}$ can be computed recursively, by iterating over all partitions $\Pi^l_{n}$ and either adding the element $n+1$ to an existing block or adding a new block containing the single element $n+1$.
The only difference is that when creating a new block, all the possible $l$ colors for this block have to be considered.

The partitions of a set into $l$ labeled blocks allow one to evaluate the following $n$th order derivative of a multivariate composite function $f[g_1(x_1,\ldots,x_n),\ldots,g_l(x_1,\ldots,x_n)]$~\cite{constantine1996multivariate,leipnik2007multivariate}:\footnote{Note that the $n$ variables $x_n$ need not be distinct.}
\eq{\label{eq:FdBmulti}
& \frac{\partial^n f[g_1(x_1,\ldots,x_n),\ldots,g_l(x_1,\ldots,x_n)]}{\partial x_1 \, \ldots \, \partial x_n} = \non
& \quad \sum_{\pi \in \Pi^l_n} 
\frac{\partial^{|\pi|} f}{\partial g_1^{|\pi_1|} \ldots \partial g_l^{|\pi_l|} }
\,
\prod_{j=1}^l  \prod_{B \in \pi_j} \frac{\partial^{|B|} g_l}{\prod_{j \in B} \partial x_j}~.
}
Here $\pi_j$ is the subset of $\pi$ which contains all blocks of color $j$, $|\pi_j|$ is the dimension of this set and $|\pi| = \sum_{j=1}^l |\pi_j|$.
It is instructive to consider a special case of $l = 2$ and $f = f[g_1(x,y),g_2(x,y)]$, which is relevant for the applications studied in the present work. The derivative $\partial^{n+m} f/ \partial x^n \partial y^m$ reads
\eq{\label{eq:FdBtwo}
& \frac{\partial^{n+m} f[g_1(x,y),g_2(x,y)]}{\partial x^n \, \partial y^m} = \non
& \quad \sum_{\pi \in \Pi^2_n} 
\frac{\partial^{|\pi_1|+|\pi_2|} f}{\partial g_1^{|\pi_1|} \partial g_2^{|\pi_2|} }
\,
\prod_{B \in \pi_1} \frac{\partial^{|B|} g_1}{\partial x^{B_1} \partial y^{B_2}}
\prod_{B \in \pi_2} \frac{\partial^{|B|} g_2}{\partial x^{B_1} \partial y^{B_2}}~.
}
Here $B_1$ is the number of elements in block $B$ that are smaller or equal than $n$ and $B_2$ is the number of all other elements of $B$.

\section{Derivatives of $f_{\inout}$ in terms of grand-canonical cumulants}
\label{app:invsingle}

The grand-canonical cumulants $\kappa_{n,\inout}^{\rm gce}$ are defined implicitly through Eqs.~\eqref{eq:maximum2} and~\eqref{eq:gcedef}:
\eq{\label{eq:eqssingle}  
f_{\inout}'[\mean{B_{\inout}}(t)] &= t, \\
\kappa_{n,\inout}^{\rm gce} & \equiv \left. \frac{\partial^{n-1} \mean{B_{\inout}}(t)}{\partial t^{n-1}} \right|_{t=0}.
}
One can obtain a recursive procedure to express the derivatives of $f$ in terms of the grand-canonical cumulants $\kappa_{n,\inout}^{\rm gce}$.  
Such a procedure has been derived in Ref.~\cite{Savchuk:2019yxl} using the FdB formula.
Evaluating the $n$th order derivative of Eq.~\eqref{eq:eqssingle} with respect to $t$ with through the FdB formula~\eqref{eq:FdBstand} one obtains
\eq{
\sum_{k=1}^n \, f_{\inout}^{(k+1)} \, B_{n,k} \left[\kappa_{2,\inout}^{\rm gce}, \ldots, \kappa_{n-k+2,\inout}^{\rm gce}\right] = \delta_{n,1}.
}
Solving the equation for $f_{\inout}^{(n+1)}$ yields,
\eq{
& f_{\inout}^{(n+1)} =  \non 
& \frac{\delta_{n,1} - \sum_{k=1}^{n-1} \, f_{\inout}^{(k+1)} \, B_{n,k} \left[\kappa_{2,\inout}^{\rm gce}, \ldots, \kappa_{n-k+2,\inout}^{\rm gce}\right]}{(\kappa_{2,\inout}^{\rm gce})^n},
}
which is a recursive relation defining $f_{\inout}^{(n+1)}$ in terms of lower order derivatives $f^{(2)},\ldots,f^{(n)}$ and the grand-canonical cumulants.
The results for $n=2,3,4$ read
\eq{
f''_{\inout} & = \frac{1}{\kappa_{2,\inout}^{\rm gce}}, \\
\label{eq:fthreegce}
f'''_{\inout} & = -\frac{f''_{\inout} \kappa_{3,\inout}^{\rm gce}}{(\kappa_{2,\inout}^{\rm gce})^2} \non
& = -\frac{\kappa_{3,\inout}^{\rm gce}}{(\kappa_{2,\inout}^{\rm gce})^3}, \\
f^{(4)}_{\inout} & = -\frac{3\,f^{(3)}_{\inout}\,\kappa_{2,\inout}^{\rm gce} \kappa_{3,\inout}^{\rm gce} + f^{(2)}_{\inout}\,\kappa_{4,\inout}^{\rm gce}}{(\kappa_{2,\inout}^{\rm gce})^3} \non
& = \frac{3 \, (\kappa_{3,\inout}^{\rm gce})^2-\kappa_{2,\inout}^{\rm gce} \kappa_{4,\inout}^{\rm gce}}{(\kappa_{2,\inout}^{\rm gce})^5}~.
}

\section{Derivatives of $g_{\inout}$ in terms of grand-canonical cumulants}
\label{app:invtwo}

The highly peaked joint distribution $W^{\rm gce}_{\inout} (B_{\inout},N_{\inout})$ of $B_{\inout}$ and $N_{\inout}$ from Sec.~\ref{sec:noncons} reads
\eq{\label{eq:Wgceapp}
W^{\rm gce}_{\inout} (B_{\inout},N_{\inout}) = C_{\inout}^{\rm gce} \, e^{-g_{\inout}(B_{\inout}, N_{\inout})}.
}
One can derive the relations between the function $g_{\inout}$ and the ``grand-canonical'' cumulants $\kappa^{\rm gce}_{\inout}$.
Introducing the cumulant generating function $G^{\rm gce}_{B_{\inout},N_{\inout}}(t_B,t_N)$ and applying the maximum term method one obtains the following equations defining $\mean{B_{\inout}}(t_B,t_N)$ and $\mean{N_{\inout}}(t_B,t_N)$:
\eq{\label{eq:tBgce}
t_B & = g_{\rm in}^{(1,0)}\left[\mean{B_{\inout}}(t_B,t_N), \mean{N_{\inout}}(t_B,t_N)\right], \\
\label{eq:tNgce}
t_N & = g_{\rm in}^{(0,1)}\left[\mean{B_{\inout}}(t_B,t_N), \mean{N_{\inout}}(t_B,t_N)\right].
}
The ``grand-canonical'' cumulants are defined as
\eq{\label{eq:kappagce1}
\kappa^{\rm gce}_{n+1,m;\inout} &= \frac{\partial^{n+m} \mean{B_{\inout}}(t_B,t_N)}{\partial (t_B)^n \partial (t_N)^m}, \\
\label{eq:kappagce2}
\kappa^{\rm gce}_{n,m+1;\inout} &= \frac{\partial^{n+m} \mean{N_{\inout}}(t_B,t_N)}{\partial (t_B)^n \partial (t_N)^m}~.
}

\begin{widetext}

Using Eqs.~\eqref{eq:tBgce} and~\eqref{eq:tNgce} one can rewrite the identities $\mean{B_{\inout}}(t_B,t_N) = \mean{B_{\inout}}$ and $\mean{N_{\inout}}(t_B,t_N) = \mean{N_{\inout}}$ as follows:
\eq{\label{eq:Bineq}
\mean{B_{\inout}}\left[g_{\inout}^{(1,0)}(\mean{B_{\inout}},\mean{N_{\inout}}),g_{\inout}^{(0,1)}(\mean{B_{\inout}},\mean{N_{\inout}})\right] & = \mean{B_{\inout}}, \\
\label{eq:Nineq}
\mean{N_{\inout}}\left[g_{\inout}^{(1,0)}(\mean{B_{\inout}},\mean{N_{\inout}}),g_{\inout}^{(0,1)}(\mean{B_{\inout}},\mean{N_{\inout}})\right] & = \mean{N_{\inout}}~. 
}

The recursive procedure to evaluate $g_{\rm in}^{(n,m)}$ for $n+m \geq 2$ in terms of $\kappa^{\rm gce}_{n,m;\inout}$ is obtained by applying a derivative $\partial^{(n+m)}/\partial \mean{B_{\inout}}^n \partial \mean{N_{\inout}}^m$ to Eqs.~\eqref{eq:Bineq},~\eqref{eq:Nineq} and evaluating it via the FdB formula~\eqref{eq:FdBtwo}:
\eq{\label{eq:g1eq}
\sum_{\{\pi_1, \pi_2\} \in \Pi_{n+m}^2}
\kappa^{\rm gce}_{|\pi_1|+1,|\pi_2|;\inout}
\prod_{B \in \pi_1}
g_{\inout}^{(1+B_1,B_2)} 
\prod_{B \in \pi_2}
g_{\inout}^{(B_1,1+B_2)}  & = \delta_{n,1} \, \delta_{m,0} ~, \\
\label{eq:g2eq}
\sum_{\{\pi_1, \pi_2\} \in \Pi_{n+m}^2}
\kappa^{\rm gce}_{|\pi_1|,|\pi_2|+1;\inout}
\prod_{B \in \pi_1}
g_{\inout}^{(1+B_1,B_2)} 
\prod_{B \in \pi_2}
g_{\inout}^{(B_1,1+B_2)}  & = \delta_{n,0} \, \delta_{m,1} ~.
}
The sum $\sum_{\{\pi_1, \pi_2\} \in \Pi_{n+m}^2}$ contains two terms which correspond to partitions of the set $\{1,\ldots,n+m\}$ into a single block, of colors 1 and 2, respectively.
Separating these terms from the sums in Eqs.~\eqref{eq:g1eq} and~\eqref{eq:g2eq} allows one to rewrite these expressions as a linear system of equations for $g_{\inout}^{(1+n,m)}$ and $g_{\inout}^{(n,1+m)}$:
\eq{\label{eq:g1leq}
\kappa^{\rm gce}_{2,0;\inout} \, g_{\inout}^{(1+n,m)} + \kappa^{\rm gce}_{1,1;\inout} \, g_{\inout}^{(n,m+1)} & =  \delta_{n,1} \, \delta_{m,0} - \sum_{\{\pi_1, \pi_2\} \in \tilde{\Pi}_{n+m}^2}
\kappa^{\rm gce}_{|\pi_1|+1,|\pi_2|;\inout}
\prod_{B \in \pi_1}
g_{\inout}^{(1+B_1,B_2)} 
\prod_{B \in \pi_2}
g_{\inout}^{(B_1,1+B_2)}, \\
\label{eq:g2leq}
\kappa^{\rm gce}_{1,1;\inout} \, g_{\inout}^{(1+n,m)} + \kappa^{\rm gce}_{0,2;\inout} \, g_{\inout}^{(n,m+1)} & =  \delta_{n,0} \, \delta_{m,1} - \sum_{\{\pi_1, \pi_2\} \in \tilde{\Pi}_{n+m}^2}
\kappa^{\rm gce}_{|\pi_1|,|\pi_2|+1;\inout}
\prod_{B \in \pi_1}
g_{\inout}^{(1+B_1,B_2)} 
\prod_{B \in \pi_2}
g_{\inout}^{(B_1,1+B_2)}~.
}
Here  $\tilde{\Pi}_{n+m}^2 = \Pi_{n+m}^2 \setminus \{|\pi_1| + |\pi_2| = 1\}$ corresponds to all partitions that contain more than a single block.
The sums in the r.h.s of Eqs.~\eqref{eq:g1leq} and~\eqref{eq:g2leq} contain the derivatives of $g$ of order up to $n+m$.
Thus, solving this linear system of equations provide a recursive relation to calculate $g_{\inout}^{(1+n,m)}$ and $g_{\inout}^{(n,1+m)}$ in terms of the cumulants $\kappa^{\rm gce}_{\inout}$. The recursive procedure starts from $n+m = 1$.

The first-order derivatives $g_{\inout}^{(1,0)}$ and $g_{\inout}^{(0,1)}$ vanish at $t_B = t_N = 0$, following Eqs.~\eqref{eq:tBgce},~\eqref{eq:tNgce}.
The second order derivatives $g_{\inout}^{(2,0)}$, $g_{\inout}^{(1,1)}$, and $g_{\inout}^{(0,2)}$ are obtained by solving the linear system of equations~\eqref{eq:g1leq} and~\eqref{eq:g2leq} for $n = 1,\, m = 0$ and $n = 0,\, m=1$:
\eq{
g_{\inout}^{(2,0)} & = \frac{\kappa^{\rm gce}_{0,2;\inout}}{ \kappa^{\rm gce}_{2,0;\inout} \kappa^{\rm gce}_{0,2;\inout} - ( \kappa^{\rm gce}_{1,1;\inout})^2}~, \\
g_{\inout}^{(1,1)} & = -\frac{\kappa^{\rm gce}_{1,1;\inout}}{ \kappa^{\rm gce}_{2,0;\inout} \kappa^{\rm gce}_{0,2;\inout} - ( \kappa^{\rm gce}_{1,1;\inout})^2}~, \\
g_{\inout}^{(0,2)} & = \frac{\kappa^{\rm gce}_{2,0;\inout}}{ \kappa^{\rm gce}_{2,0;\inout} \kappa^{\rm gce}_{0,2;\inout} - ( \kappa^{\rm gce}_{1,1;\inout})^2}~.
}

The higher order derivatives of $g$ are obtained by following the developed recursive procedure.

\end{widetext}

\bibliography{SAM2p0}


\end{document}